\documentclass[epj]{svjour}
\usepackage{graphics}
\begin{document}
\title{Evidence of Final-State Suppression of High-p$_{_{\rm T}}$
Hadrons in Au~+~Au~Collisions Using d~+~Au Measurements at RHIC}
\author{Rachid Nouicer$^{2,6}$ for the PHOBOS Collaboration \\[2mm]
B.B.Back$^1$,
M.D.Baker$^2$,
M.Ballintijn$^4$,
D.S.Barton$^2$,
B.Becker$^2$,
R.R.Betts$^6$,
A.A.Bickley$^7$,
R.Bindel$^7$,
W.Busza$^4$,
A.Carroll$^2$,
M.P.Decowski$^4$,
E.Garc\'{\i}a$^6$,
T.Gburek$^3$,
N.George$^2$,
K.Gulbrandsen$^4$,
S.Gushue$^2$,
C.Halliwell$^6$,
J.Hamblen$^8$,
A.S.Harrington$^8$,
C.Henderson$^4$,
D.J.Hofman$^6$,
R.S.Hollis$^6$,
R.Ho\l y\'{n}ski$^3$,
B.Holzman$^2$,
A.Iordanova$^6$,
E.Johnson$^8$,
J.L.Kane$^4$,
N.Khan$^8$,
P.Kulinich$^4$,
C.M.Kuo$^5$,
J.W.Lee$^4$,
W.T.Lin$^5$,
S.Manly$^8$,
A.C.Mignerey$^7$,
R.Nouicer$^{2,6}$,
A.Olszewski$^3$,
R.Pak$^2$,
I.C.Park$^8$,
H.Pernegger$^4$,
C.Reed$^4$,
C.Roland$^4$,
G.Roland$^4$,
J.Sagerer$^6$,
P.Sarin$^4$,
I.Sedykh$^2$,
W.Skulski$^8$,
C.E.Smith$^6$,
P.Steinberg$^2$,
G.S.F.Stephans$^4$,
A.Sukhanov$^2$,
M.B.Tonjes$^7$,
A.Trzupek$^3$,
C.Vale$^4$,
G.J.van~Nieuwenhuizen$^4$,
R.Verdier$^4$,
G.I.Veres$^4$,
F.L.H.Wolfs$^8$,
B.Wosiek$^3$,
K.Wo\'{z}niak$^3$,
B.Wys\l ouch$^4$,
J.Zhang$^4$\\
}
\institute{  
$^1$~Argonne National Laboratory, Argonne, IL 60439-4843, USA\\
$^2$~Brookhaven National Laboratory, Upton, NY 11973-5000, USA\\
$^3$~Institute of Nuclear Physics, Krak\'{o}w, Poland\\
$^4$~Massachusetts Institute of Technology, Cambridge, MA 02139-4307, USA\\
$^5$~National Central University, Chung-Li, Taiwan\\
$^6$~University of Illinois at Chicago, Chicago, IL 60607-7059, USA\\
$^7$~University of Maryland, College Park, MD 20742, USA\\
$^8$~University of Rochester, Rochester, NY 14627, USA
}
\vspace{-1cm}
\date{Received: date / Revised version: date}
\abstract{Transverse momentum spectra of charged hadrons with 
${p_{T} <}$ 6~GeV/c have been measured near mid-rapidity (0.2 $< \eta <$
1.4) by the PHOBOS experiment at
RHIC in Au $+$ Au and d $+$ Au collisions at ${\sqrt{s_{_{NN}}}~= \rm {200~GeV}}$. 
The spectra for different collision
centralities are compared to ${p + \bar{p}}$ collisions at
the same energy. The
resulting nuclear modification factor for central Au $+$ Au
collisions shows evidence of strong suppression of charged hadrons in the high-$p_{T}$
region (${>2}$ GeV/c). 
In contrast, the d $+$ Au nuclear modification factor exhibits
no suppression of the high-$p_{T}$ yields. These measurements suggest
a large energy loss of the high-$p_{T}$ particles in the highly
interacting medium created in the central Au $+$ Au collisions. 
The lack of suppression in d $+$ Au collisions suggests that it is unlikely that
initial state effects can explain the suppression in the central Au
$+$ Au collisions.  
\PACS{ 25.75.-q
%      {PACS-key}{describing text of that key}   \and
%      {PACS-key}{describing text of that key}
     } % end of PACS codes
} %end of abstract
\authorrunning{R. Nouicer}
\titlerunning {Evidence of Final-State Suppression of High-p$_{_{\rm
T}}$ Hadrons in~Au~+~Au~Collisions Using d~+~Au...  }
\maketitle
\section{Introduction}
\label{intro}
\vspace*{-0.3cm}
In the theoretical analysis of particle production 
in hadronic and nuclear collisions, a distinction is often made
between the relative contributions from ``hard'' parton-parton 
scattering processes and ``soft'' processes. 
The contribution from hard processes is expected to grow with
increasing energy and to dominate particle production at high
transverse momentum. 
Collisions of heavy nuclei offer ideal conditions to test our
understanding of this picture, as ``hard'' processes are expected to
scale with the number of binary nucleon-nucleon collisions ${\rm
N_{\rm coll}}$, whereas ``soft'' particle production is expected to
exhibit scaling with the number of participant nucleons 
${\rm N_{\rm part}}$. In Glauber-model calculations, N$_{\rm coll}$ scales 
approximately as (N$_{\rm part}$)$^{4/3}$. For Au $+$ Au collisions at
the Relativistic Heavy Ion Collider (RHIC) energies, it has been
predicted that the yield and momentum distribution of particles
produced by hard scattering processes may be modified by ``jet
quenching'', i.e. the energy loss of high momentum partons in the
dense medium~\cite{Gyl1,Gyl2}. 
\par
The data for Au $+$ Au collisions at ${ \sqrt{s_{_{NN}}}~ = \rm
{200~GeV} }$ were collected using the PHOBOS two-arm magnetic 
spectrometer~\cite{NIM} at RHIC. The spectrometer arms are each equipped 
with 16 layers of silicon sensors, providing charged particle
tracking both outside and inside 
the 2T field of the PHOBOS magnet. The primary event trigger and event
selection were
provided by two sets of 16 scintillator counters.
For d $+$ Au collisions at ${ \sqrt{s_{_{NN}}}~ = \rm {200~GeV} }$ 
an additional array of silicon detectors was used in the event selection. 
The array consisted of the central
single-layer Octagon barrel detector and the three single-layer
forward Ring detectors located on either side of the interaction
point. In addition, two higher level trigger conditions were used
which utilized two rings of ten Cerenkov counters around the beam
pipe and two arrays 
of horizontally segmented scintillator hodoscopes behind the spectrometer. 
The yields of charged hadrons produced in Au + Au and d + Au collisions
at ${ \sqrt{s_{_{NN}}}~ = \rm {200~GeV} }$ 
are presented as a function of collision 
centrality and transverse momentum~${p_{T}}$.
\vspace*{-0.7cm}
\section{Observation of suppression of high-p$_{_{\rm T}}$ particles in
Au + Au at {\bf ${ \sqrt{s_{_{NN}}}~ = \rm {200~GeV} }$ }}
\label{sec:1}
\begin{figure}[hb]
\begin{center}
\hspace*{-0.2cm}\resizebox{0.5\textwidth}{!}{
\includegraphics{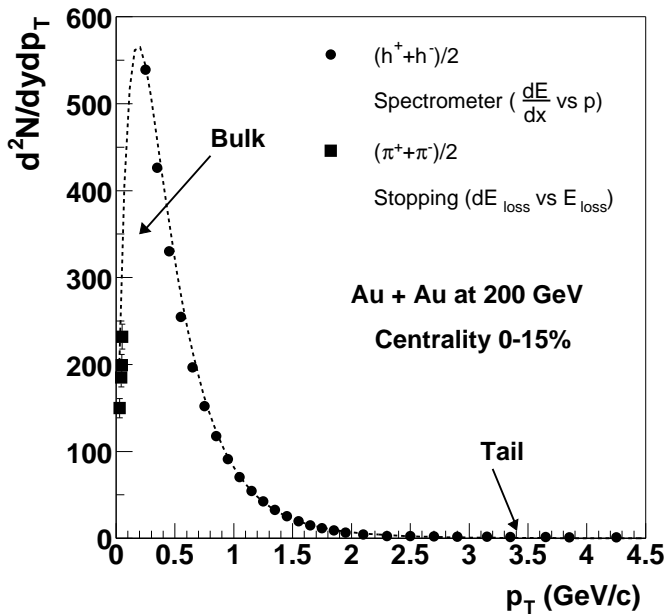}
}
\caption{Charged hadron distribution as a function of
transverse momentum ${ p_{T}}$ for the 0--15$\%$ most central Au + Au collisions at 
${ \sqrt{s_{_{NN}}}~ = \rm {200~GeV} }$. 
The systematic errors are not shown.}
\label{fig:1}      
\end{center}
\end{figure}

The charged hadron distribution produced in the 0--15$\%$ most central 
Au $+$ Au collisions at 
${ \sqrt{s_{_{NN}}}~ = \rm {200~GeV} }$ is presented in Fig.~\ref{fig:1}. The
distribution illustrates a ``bulk'' and ``tail'' which can be related
to ``soft'' processes and ``hard'' parton-parton scattering
respectively. It should be kept in mind, however, that there is no 
clear separation between ``hard'' and soft processes. For this reason,
an analysis as a function of centrality and transverse momentum is
required to better understand particle production. 
\vspace*{-0.7cm}
\par
\begin{figure}[hb]
\begin{center}
\hspace*{-0.1cm}\resizebox{0.43\textwidth}{!}{
\includegraphics{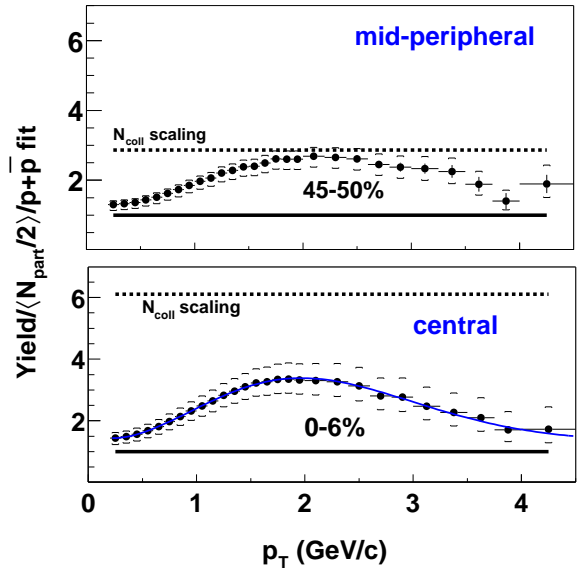}
}
\caption{Ratio of the yield of charged hadrons as a function of ${
p_{_{\rm T}}}$ for the most peripheral bin (${\rm \langle N_{part} \rangle = 65 \pm
4}$, upper plot) and the most central (${\rm \langle N_{part} \rangle = 344 \pm
11}$, lower plot) scaled by ${\rm \langle N_{part} \rangle /2}$ and
normalized to the fit to proton-antiproton data.
The dashed (solid) line shows the expectation
of ${\rm N_{coll} (N_{part})}$ scaling relative to ${p + \bar{p}}$
collisions. The brackets show the systematic uncertainty of the Au $+$
Au data.
}
\label{fig:2}      
\end{center}
\end{figure}

To study the evolution of the spectra with centrality in detail, 
we divide the data by a fit to the $p_{T}$-distribution
measured in proton-antiproton
collisions at the same energy~\cite{proton}. The ${p + \bar{p}}$ data
were translated into the PHOBOS acceptance using PYTHIA, following the
procedure described in Ref.~\cite{method}. 
Fig.~\ref{fig:2} shows the Au $+$ Au yields divided by ${\rm \langle N_{part}
\rangle /2}$ for the mid-peripheral
bin (${\rm \langle N_{part} \rangle = 65 \pm 4}$) and the most central
bin  (${\rm \langle N_{part} \rangle = 344 \pm 11}$) scaled by the fit to the
measured ${p_{T}}$-distributions in proton-antiproton
collisions. The brackets indicate the
systematic uncertainty in the Au $+$ Au data (90$\%$  C.L.). The
largest contributions to the systematic uncertainty are the overall
tracking efficiency, independent of ${p_{T} }$, and the 
${p_{T} }$-dependent momentum resolution and binning
correction. 
The proximity of the tracking to the collision vertex ensures that the 
contamination by secondary particles and feed-down particles is small. 
Therefore the correction factor and the uncertainty are also small.
Similarly, the high
granularity and resolution of the tracking planes leads to a small
uncertainty in the rate of ``ghost'' tracks.
\vspace*{-0.5cm}  
\vspace*{-0.9cm}
\par 
As has been shown previously~\cite{pr1,pr2}, the yield per participant
pair in Au $+$ Au collisions at these centralities is significantly
larger than in proton-antiproton collisions at the same energy. We
also observe that in the mid-peripheral Au $+$ Au collisions with 
${\rm \langle N_{part} \rangle = 65}$ (corresponding to impact parameter 
${ b \sim 10 }$~fm), the spectral shape is already strongly modified from
that in ${ p + \bar{p} }$ collisions. It is worth noting that the
ratio ${\rm \frac{\langle N_{coll} \rangle}{\langle N_{part}/2 \rangle}}$ 
increases by a factor of almost three from ${ p + \bar{p} }$ to the 
mid-peripheral Au $+$ Au collisions studied here. For the highest
${p_{\rm T} }$, the yield for central events is significantly
smaller than expectations based on ${\rm N_{ coll}}$-scaling. 
\vspace*{-0.5cm}
\section{Absence of suppression of high-p$_{_{\rm T}}$ particles in
d + Au at {\bf ${ \sqrt{s_{_{NN}}}~ = \rm {200~GeV} }$ }}
\label{sec:3}
The d $+$ Au measurement at higher energies is motivated by results
from Au $+$ Au collisions at ${ \sqrt{s_{_{NN}}}~}$ = 130 and
200~GeV. In these collisions, the expected scaling of hadron
production with the number of binary nucleon-nucleon collisions at 
${p_{\rm T}}$ from 2 to 10 GeV/c is strongly
violated~\cite{Adc1,Adl,Bac1}. This effect had been predicted 
as a consequence of the energy loss of high-$p_{T}$ partons in the 
hot and dense medium formed in Au $+$ Au collisions~\cite{phe}. 
The interpretation of the Au $+$ Au data relies on an understanding
of initial state effects, including gluon saturation~\cite{sat}, which
can be investigated with the d $+$ Au data presented here. By studying
the spectra as a function of collision centrality, we can control the
effective thickness of nuclear matter traversed by the incoming
partons.
\vspace*{-0.5cm}\begin{figure}[hbt]
\hspace*{-0.5cm}\resizebox{0.5\textwidth}{!}{
\includegraphics{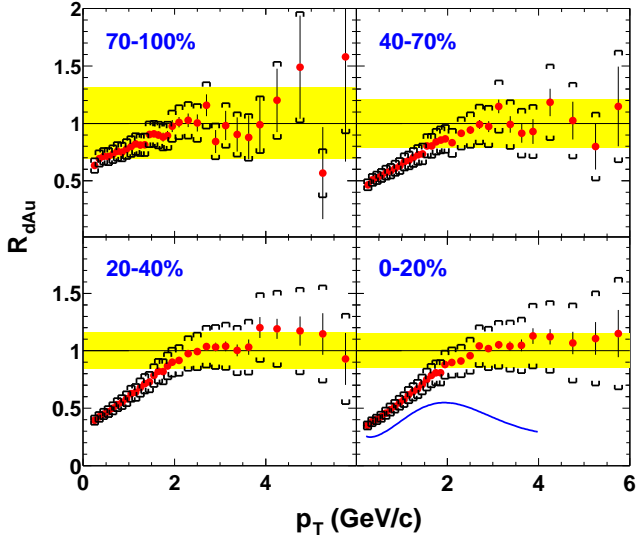}
}
\caption{Nuclear modification factor ${R_{dAu}}$ for d $+$ Au
collisions as a function of
${p_{T}}$ for four centrality bins. In the most central
bin (0--20$\%$), the spectral shape for central Au $+$ Au data relative to
${p + \bar{p} }$ (UA1) is shown for comparison as a solid line. The shaded
area shows the uncertainty in ${R_{dAu}}$ due to the systematic
uncertainty in ${\rm \langle N_{coll} \rangle}$ and the UA1 scale
error (90$\%$ C.L.). The brackets show the systematic uncertainty of
the d $+$ Au spectra measurement (90$\%$ C.L.).  
  }
\label{fig:3}      
\end{figure}
\vspace*{-0.5cm}
\par
In Fig.~\ref{fig:3} we present the nuclear modification factor
$R_{dAu}$ as a function of ${p_{T}}$ for each centrality bin,
defined as: 
$${
R_{dAu} =\frac{ \sigma^{inel}_{p\bar{p}} d^{2} N_{dAu}/dp_{T}d\eta}{
\langle N_{coll} \rangle d^{2}\sigma({\rm UA1})_{p\bar{p}}/dp_{T}d\eta} 
}$$
Consistent with our Glauber calculation, we used
${\sigma^{inel}_{p\bar{p}}}$ = 41 mb. A value of ${R_{dAu} =1}$
corresponds to scaling of the yield as an incoherent superposition 
of nucleon-nucleon collisions. For all centrality bins, we observe 
rapid rises of $R_{dAu}$ from low $p_{T}$, leveling off at 
${p_{T}}$ of $\sim$ 2 GeV/c. For comparison, we also plot
the results from central Au $+$ Au collisions at the same
energy~\cite{Bac1} in the lower right panel of Fig.~\ref{fig:3}
as a solid line. 
The average number of collisions undergone by each participating
nucleon in the central Au $+$ Au collision is close to 6, similar to
that of each nucleon from the deuteron in a central d$+$Au
collision. For central Au $+$ Au collisions, the ratio of the spectra
to ${p + \bar{p}}$ rises rapidly up to $p_{T}=$ 2 GeV/c, but falls
far short of collision scaling at larger $p_{T}$, in striking contrast
to the behavior for central d $+$ Au collisions. 
Predictions for the evolution of $R_{dAu}$ from semi-peripheral
collisions with ${\rm \langle N_{coll} \rangle \sim 6 }$ to central
collisions were made in two qualitatively different
models.
%\vspace*{-0.2cm} 
\begin{figure}[htb]
\begin{center}
\resizebox{0.35\textwidth}{!}{
\includegraphics{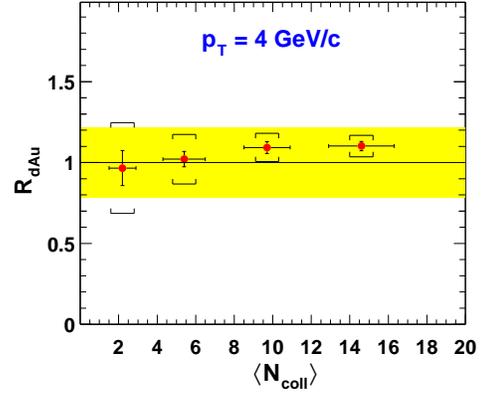}
}
\caption{Nuclear modification factor for R$_{dAu}$ as a function of
centrality at $p_{T}$ = 4 GeV/c. The brackets
indicate the point-to-point systematic error, dominated by the
uncertainty in the number of collisions for each centrality bin. The
shaded band shows the overall scale uncertainty.
Systematic errors are at 90$\%$ C.L.
}
\label{fig:4}      
\end{center}
\end{figure}
\vspace*{-0.2cm}
%\newline
\par
The results of a perturbative QCD calculation~\cite{QCD}
predict an increase in the maximum value of $R_{dAu}$ at $p_{T} \sim
3.5$ GeV/c of 15$\%$. In contrast, a decrease in $R_{dAu}$ by
25--30$\%$ over the same centrality range is predicted in the parton
saturation model~\cite{sat}. The centrality evolution of $R_{dAu}$ 
is shown in Fig~\ref{fig:4} for transverse momentum ${p_{T} = }$~4
GeV/c, where the points were obtained from a fit
to the $p_{T}$ dependence of $R_{dAu}$ in each centrality bin. 
We extract the ratio ${\frac{R_{dAu} (N_{coll} = 14.6)}{R_{dAu}
(N_{coll}=5.4)} = 1.08 \pm 0.06 }$ (the systematic uncertainty
corresponds to 15--20$\%$ at 90$\%$ C.L.).  Our data therefore
disfavor the prediction from the parton saturation model. 
The lack of suppression in d $+$ Au collisions suggests that it is unlikely that
initial state effects can explain the suppression observed in central Au
$+$ Au collisions.\\  
\par
This work was partially supported by U.S. DOE grants DE-AC02-98CH10886,
DE-FG02-93ER40802, DE-FC02-94ER40818, DE-FG02-94ER40865, DE-FG02-99ER41099,
and W-31-109-ENG-38, US NSF grants 9603486, 9722606 and 0072204, Polish KBN
grant 2-P03B-10323, and NSC of Taiwan contract NSC 89-2112-M-008-024.

\end{document}